\documentclass[a4paper,10pt,pra,twocolumn]{revtex4}
\usepackage[english]{babel}
\usepackage[utf8]{inputenc}
\usepackage{amsmath}
\usepackage{setspace}
\usepackage{array}
\usepackage{times}
\usepackage{amssymb}
\usepackage{amstext}
\usepackage{amsfonts}
\usepackage{mathrsfs}
\usepackage[all]{xy}
\usepackage{pgf}
\usepackage{svgcolor}
\usepackage{tikz}
\usepackage{rotating}
\usepackage{amsthm}
\usepackage{version}
\usepackage{placeins}

\newtheorem{ppt}{Proposition}

\newcommand{\ket}[1]{|#1\rangle}
\newcommand{\bra}[1]{\langle#1|}

\newcommand{\C}{\mathbb {C}}

\begin{document}
\begin{abstract}

We present a general scheme for sharing quantum secrets, and an extension to sharing classical secrets, which contain all known quantum secret sharing schemes. In this framework we show the equivalence of existence of both schemes, that is, the existence of a scheme sharing a quantum secret implies the extended classical secret sharing scheme works, and vice versa. As a consequence of this we find new schemes sharing classical secrets for arbitrary access structures. We then clarify the relationship to quantum error correction and observe several restrictions thereby imposed, which for example indicates that for pure state threshold schemes the share size $q$ must scale with the number of players $n$ as $q\geq \sqrt{n}$. These results also provide a new way of searching for quantum error correcting codes.

\end{abstract}

\title{On the equivalence between sharing quantum and classical secrets, and error correction}
\author{Anne Marin}
\email[]{anne.marin@telecom-paristech.fr}
\author{Damian Markham}
\email[]{damian.markham@telecom-paristech.fr}
\affiliation{CNRS LTCI, D\'{e}partement Informatique et R\'{e}seaux, Telecom ParisTech, 23 avenue d'Italie, CS 51327,  75214 Paris CEDEX 13, France.}
\maketitle

Secret sharing~\cite{Sha79} is an important primitive in information networks, for example in online auctions, electronic voting, secure multiparty function evaluation. The problem setting is that a dealer $d$ wishes to distribute a secret (we will consider both classical and quantum secrets) to a set of $n$ players, such that only certain sets of players can access the secret (we call these the \textit{authorised} sets of players). The sets that do not have access to any information about the secret are called the \emph{unauthorised} sets. The assignement of authorised sets is the \emph{access structure}. 
Any such scheme can be loosely described as a \emph{ramp} scheme, written in terms of three parameters, $(k,k',n)$, where any set of players $B$ such that $|B|\geq k$ can access the secret, whereas any set such that $|B|\leq k'$ cannot get any information at all. Clearly in general this description does not cover the full access structure in between $k$ and $k'$. When $k'=k-1$, it does however, and this is called a perfect threshold scheme, denoted $(k,n)$, perfect because a subset is either authorised or unauthorised, and threshold because no subset of cardinality less than $k-1$ is authorised. In this work when we refer to threshold schemes we assume perfect schemes also, although this is not an assumption always made in the literature (e.g. \cite{ISN89}). Often it suffices to consider threshold schemes $(k,n)$, since all access structure can be built from them (\cite{ISN89,Gottesman00}), although the efficiency of such schemes is not always optimal (\cite{SR10}).

We consider two quantum extensions of the secret sharing problem, first put forward in \cite{HBB99,CGL99}, which have found application, for example, in secure multiparty quantum computation \cite{BCGHS06}. The first is the sharing of a quantum secret \cite{CGL99,HBB99}, that is, the dealer wishes to distribute a quantum state such that only authorised sets of players can access it, and unauthorised sets cannot. We refer to this protocol family as QQ (following the notation of \cite{MS08}). It was shown in \cite{CGL99} that all threshold not contradicting no-cloning can be achieved. Here we will see that whilst this is true, error correction implies severe restrictions on how this can be done in particular in the dimension of the systems used.
The second quantum version we consider is the sharing of a classical secret using quantum channels, introduced in \cite{HBB99}. This family of protocols is referred to as CQ \cite{MS08}. It is known that there exist informationally theoretically secure schemes to share a classical secret \cite{Sha79}, however, these schemes require a secure channel between the dealer and each player. One way of resolving this issue would be to use $n$ quantum key distribution (QKD) channels from the dealer, one to each player, and then use the Shamir scheme. Another way, presented in \cite{HBB99} combines the idea of QKD with secret sharing directly. By choosing a suitable entangled state shared between the dealer and the players, the dealer is able to share a private (secure) key with the players such that only authorised players can access the key. We refer to protocols taking this second approach as RCQ (transmission of a Random Classical key with a multiparty Quantum state or channel). The existing RCQ schemes to date are threshold schemes with parameters $(n,n)$\cite{HBB99}, $(3,5)$ \cite{MS08} and $(2,3)$ \cite{Keet10} \cite{NOTE}. Although these RCQ schemes may be less practical because of the entanglement than the simple QKD schemes, we study them here for two main reasons. First, we believe these schemes are of intrinsic interest, with potential as building blocks of more elaborate protocols, and moreover second, through the relationship we present they can be useful to search for new QQ and error correcting schemes. We finally remark that we are not considering explicitly the CC protocols of \cite{MS08}, corresponding to the simple sharing of a classical secret with a quantum state, though many connections to QQ can be carried through the equivalence to RCQ.

In \cite{MS08,Keet10} a link was presented between QQ and RCQ protocols where it was shown that in some instances the same framework could be used for both using graph states. The usefulness of this connection is many fold. On a practical level sharing the same framework is advantageous since any implementation for one can be adapted to perform the other. On the theoretical level the advantages are very rich. On the one hand it allows new RCQ schemes to be found via translation from QQ, as in \cite{MS08,Keet10}. In the other direction, techniques for constructing RCQ schemes (which can often on the face of it appear much simpler) can be used to construct QQ schemes. Furthermore there is a deep relationship between QQ and quantum error correction. For example, it was shown in \cite{Sar12} that for qubit system, there is a (restricted) equivalence between QQ protocols based on graph states and CSS stabilizer codes \cite{CGL99,RST05}. This opens up the door to the possibility of using powerful tools from error correction theory to investigate secret sharing, and techniques from secret sharing to find error correcting codes. Graph state methods have recently yielded many results in this direction \cite{JMS11,Gravier11,Javelle12}. The main topic of this work is to give a general relationship between between error correction, QQ and RCQ secret sharing, which will complete previous results. We will develop this connection to show equivalence of schemes, which will lead to new RCQ schemes, tighter security of RCQ schemes, and bounds on what QQ schemes are possible, as well as providing a different approach to searching for error correcting codes.

 In this work we present the most general QQ secret sharing scheme for sharing a quantum secret and its extension to a RCQ scheme, which formally encompass all quantum existing secret sharing schemes. We show that the existence of such a QQ implies the extension to RCQ case has the same access structure, and similarly, if there exists a RCQ scheme of this type, the quantum version is also a valid QQ secret sharing scheme (propositions 1 and 2). This allows us to use the QQ schemes from \cite{CGL99} for the extended RCQ scheme, allowing all access structures (not violating the no cloning theorem). The equivalence further allows for security amongst authorised sets for the RCQ scheme.  We then clarify the equivalence between the secret sharing protocols and quantum error correction, showing that all ramp schemes are error correcting schemes and vice versa. Several restrictions are thus imposed from the theory of error correction, notably, that for pure state QQ threshold schemes, the size of the share must scale with the size of the network (something which is also true in the fully classical setting).

We start in the next section by describing the protocols. Then in section \ref{SEC: access} we give precise definitions of the requirements of authorised and unauthorised sets in terms of the information that can obtain. In section \ref{SEC: equivalence} we show how the QQ and RCQ schemes are related to each other, summarized in table I. Then in section \ref{SEC: QECC} we elaborate on the relationship with error correction and finish with discussions in section \ref{SEC: discussion}.

\bigskip
\section{QQ and RCQ secret sharing protocols} \label{SEC: protocols}

The most general QQ quantum secret sharing protocol can be understood as a map from a quantum secret state of dimension $q$, $|\zeta\rangle=\sum_{i=0}^{q-1}\alpha_i |i\rangle$ to a multipartite state $|\zeta_L\rangle_{1...n}=\sum_{i=0}^{q-1}\alpha_i |i_L\rangle_{1...n}$,
shared between the players $1...n$,  encoded onto some logical basis $\{|i_L\rangle_{1...n}\}$, which is designed such that authorised sets of players can access the secret and unauthorised sets of players cannot. Without loss of generality we take $\{|i_L\rangle\}$ to be an orthonormal basis. An encoding onto a non-orthogonal basis can be understood as some preprocessing taking the state input $|\zeta\rangle$ to a state $|\zeta'\rangle$ corresponding to the non-orthogonal encoding, and then following the map above. A mixed state encoding can always be purified into a map as above, followed by tracing out of some systems (in which case the number of active players would be less than $n$, we discuss such examples and their relation to our results in section \ref{SEC: QECC}). In this way this encoding formally represents the most general scheme.

For any such scheme, we extend to RCQ sharing classical secrets by introducing what we call a channel state between system $d$ held by the dealer and the players' systems $1...n$, which can be thought of as the dealer preparing a maximally entangled state and sending half through the encoding above,
\begin{align} \label{Eqn: Channel State}
|CS\rangle_{d,1...n}:= \frac{1}{\sqrt{q}}\sum_{i=0}^{q-1} |i\rangle_d  |i_L\rangle_{1...n}.
\end{align}
This is a maximally entangled state between $d$ and the players, and can be understood as a channel from the dealer to the players. In the QQ case this channel is used to teleport the secret from the dealer's qubit to the players (that is, it acts simply as an encoding process for the most general scheme). In the RCQ case this channel is used to establish a secure random key between the dealer and the players (as per the Ekert quantum key distribution protocol \cite{Eke91}).  In both cases it is the choice of the logical basis $\{|i_L\rangle_{1...n}\}$ gives rise to the access structure. The RCQ extension we present using this idea covers all known RCQ schemes \cite{HBB99,MS08,Keet10} (up to possible reordering of public communication steps, e.g. \cite{KKI99}).

For the RCQ extension it will be useful to define generalized Pauli operators, that are for prime dimension $q$,
$X|i\rangle = |i+1\rangle$, $Z|i\rangle = \omega^i |i\rangle$, where $\omega = e^{i 2\pi / q}$, and $q$ is the dimension of the secret. For the moment we will consider prime dimensional case and later we will see how the results also work for non-prime dimension. We further denote $|i(t)\rangle$ as the eigenstates of $X^tZ$ for $t \in \{0,...,q-1\}$ and as the eigenstate of $X$ for $t=q$. The channel state can then be expanded as
\begin{align}
|CS\rangle_{d,1...n}=\frac{1}{\sqrt{q}}\sum_{i=0}^{q-1} |i(t)\rangle_d  |i(t)_L\rangle_{1...n}, \nonumber
\end{align}
where the bases $\{|i(t)_L\rangle\}$ are also orthonormal and complementary, that is $|\langle i(t)_L|j(t')_L\rangle|^2=1/q$ when $t\neq t'$.

We will first describe the protocols, then make precise what we mean exactly by authorised and unauthorised sets for both QQ and RCQ, and security for the RCQ protocol in section \ref{SEC: access}.

\bigskip

\textbf{QQ Protocol:}
Let $|\zeta\rangle_{d'}= \sum_{i=0}^{q-1}\alpha_i|i\rangle_{d'} \in \C^q$ be the secret state in possession of the dealer.

\begin{enumerate}
\item The dealer prepares a channel state (\ref{Eqn: Channel State}), then does an extended Bell measurement over $d$ and $d'$ and appropriate corrections, leaving the state of the $n$ qudits as
\begin{equation}
|\zeta_L\rangle_{1...n} = \sum_{i=0}^{q-1} \alpha_i |i_L\rangle_{1...n}.
\label{Eqn: Q secret graph state}
\end{equation}
\item The dealer sends qudit $\ell$ of the resultant state to player $\ell$.
\item Players in authorised set $B$ follow a prescribed decoding operation $\Gamma_B$.
\end{enumerate}
The protocol is then defined by encoding basis $\{|i_L\rangle_{1...n}\}$, and decoding operations $\Gamma_B$ for each authorised set $B$. More concretely,  $\Gamma_B$ maps the reduced density matrix $\rho_B^\zeta = Tr_{V/B}(|\zeta_L\rangle\langle \zeta_L|)$ on the systems of $B$, onto the secret state $|\zeta\rangle_{b'}$ on some system $b'$. $\Gamma_B$ may be a global operation over systems $B$ (plus possible ancilla systems) and $b'$ may in $B$, or some ancilla (see Fig.~\ref{FIG: QQ}).

\textbf{RCQ Protocol:}
The RCQ protocol does not directly distribute a secret classical message from the dealer to the players, rather it is a protocol to establish a secure key between the dealer and the players, such that only authorised sets of players can access the key. In this sense it may be considered more accurately as secret key sharing. This key can then be used by the dealer to share a secret message such that it can only be read by authorised sets of players. The RCQ protocol is an extension of those presented in \cite{HBB99,MS08,Keet10}, to the more general case not necessarily using graph states. We now outline the protocol.

\begin{enumerate}
\item The dealer prepares a channel state (\ref{Eqn: Channel State}) and sends qudit $\ell$ to player $\ell$.
\item The dealer randomly chooses a $t \in \{0,...,q\}$ and measures qudit $d$ among the bases: $\{X^tZ\}_{t=0}^{q-1}$ for $t \in \{0,...,q-1\}$ or $X$ if $t=q$. We denote the result $r(t)$. The state of the players is then projected to
\begin{equation}
|r(t)_L\rangle_{1...n}.
\end{equation}
\item \textit{An authorised set} $B$ randomly measures in one of the prescribed measurements  $\{M_B^{t'}\}_{t'=0}^{q}$, with result denoted $s(t')$.
\item Repeat step $1.$ $2.$ $3.$ $m\rightarrow\infty$ times. The list of measurement results $r(t)$ and $s(t')$ are the raw keys of the dealer and players $B$ respectively.
\item \textsc{security test}: Follow standard QKD security steps (see e.g. \cite{SS10}). Through public discussion between $d$ and $B$ first sift the key followed by standard error correction and privacy amplification to generate a secure key.
\end{enumerate}
The protocol is defined by encoding basis $\{|i_L\rangle_{1...n}\}$, and measurements $\{M_B^{t'}\}_{t'=0}^{q}$ for each authorised set $B$. At the end, if the protocol is not aborted during the security step, the dealer and the authorised set share a secure key which can be used to distribute a classical secret securely. \\

\bigskip

\section{Authorised and unauthorised sets and security} \label{SEC: access}
\begin{figure}[h]
{\resizebox{!}{2.7cm}{\includegraphics{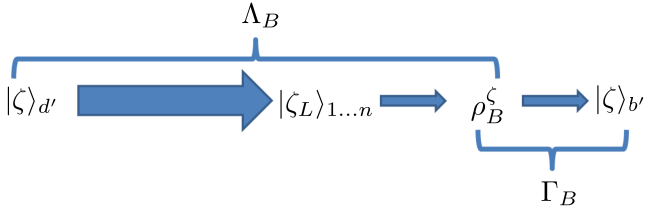}}} \caption{\label{FIG: QQ} (Color online)Schematic of the QQ scheme. A dealer encodes a secret state $|\zeta\rangle_{d'}$ onto $n$ parties (\ref{Eqn: Q secret graph state}). After tracing out of other systems to get $\rho_B^{\zeta_L}$ (together with the encoding denoted by map $\Lambda_B$), authorised players $B$ perform map $\Gamma_B$ to recover the secret.}
\end{figure}

We now define what it means to say sets of players are authorised or unauthorised for both RCQ and QQ protocols. For later proofs comparing the two protocols it will be useful to also talk about equivalent information theoretic conditions. For this we define the channel $\Lambda_B$ from system $d'$ to subset of players $B$ as the encoding procedure in QQ giving state (\ref{Eqn: Q secret graph state}) followed by tracing out all but the players $B$ (see Fig.~\ref{FIG: QQ}).

We first look at the QQ case.\\
{\bf QQ Authorised sets} We say a set of players $B$ is {\it authorised} if they can perfectly access the quantum secret, that is, if there exists a decoding procedure $\Gamma_B$ acting only on those players, which can perfectly recover the secret input state $|\zeta\rangle$.

 If the quantum information is accessible through the channel $\Lambda_B$, then the quantum mutual information between two halves of a maximally entangled state after one half has been sent down the channel is maximal \cite{SN99}. That is to say $I(\tau;\Lambda_B)=2\log_2q$, where $I(\tau;\Lambda_B)=S(\tau)+S(\Lambda_B(\tau))-S((id\otimes\Lambda_B)(\ket{\Phi_q}\bra{\Phi_q}))$, $\tau=\frac{1}{q}\sum_{i=0}^{q-1}\ket{i}\bra{i}$ is a maximally mixed state, $\ket{\Phi_q}=\frac{1}{q}\sum_{i=0}^{q-1}\ket{ii}$ is a maximally entangled state and $S$ is the Von Neuman entropy ($S(\rho)=-tr(\rho\log(\rho))$).\\
\\
{\bf QQ Unauthorised sets} We say a set of players $B$ is {\it unauthorised} if it has no access to the quantum secret whatsoever, that is, the reduced density matrix $\rho_B$ is independent of the quantum input $|\zeta\rangle$.
Information theoretically, this corresponds to $I(\tau;\Lambda_B)=0$.

\bigskip

We now look at the RCQ protocol.\\
{\bf RCQ Authorised sets} We say a set of players $B$ is \textit{authorised} if it can access the secret, that is, if after the dealer has distributed the channel state (\ref{Eqn: Channel State}) and measured it (up to step 2 in the protocol), there exists a (possibly joint) measurement on their systems which allows them to discover the dealers measurement result $r(t)$ for each setting $t$.

To rewrite this in information theoretic language, it suffices to consider the channel from the dealer to the players $B$, where for each $t$, the dealer sends a specific chosen state $|r(t)_L\rangle_{1...n}$ to the players encoding the classical information $r(t)$, chosen according to a uniform distribution. The ability, or not, of a set of players to access this classical information is equivalent to them being able to discover the dealer's measurement result in the RCQ protocol. In terms of the action of the channel $\Lambda_B$ above, this corresponds to a set of inputs $\{U^t\ket{i}\}_{i=0}^{q-1}$, where $U$is a fourier transform of rank $q$, $t\in [q]$. That is, each $|r(t)_L\rangle_{1...n}$, corresponds to an input state $U^t\ket{r}_{d'}$. Thus to verify that this channel works perfectly for each such message, we are interested in the classical information that can be transmitted for a random distribution over the alphabet for a given $t$, which we denote $\mathcal{E}_t=\{\frac{1}{q},U^t\ket{i}\}_{i=0}^{q-1}$. We use Holevo information defined over a quantum channel $\Lambda_B$ by :
{\small{$$\chi(\Lambda_B(\mathcal{E}_t))=S\Big(\frac{1}{q}\sum_{i}\Lambda_B(U^t\ket{i}\bra{i}U^{t\dagger})\Big)-\frac{1}{q}\sum_{i}S(\Lambda_B(U^t\ket{i}\bra{i}U^{t\dagger})).$$}}
If the classical information is accessible through the quantum channel $\Lambda_B$ perfectly, then the Holevo information $\chi(\mathcal{E}_t(\Lambda_B))=\log q$, which must be true for all $t$, for all authorised sets $B$.\\
\\
{\bf RCQ Unauthorised sets} We say a set of players $B$ is {\it unauthorised} if the dealer's result $r(t)$ is completely denied to them, that is, if the reduced state $\rho_B$ of those systems has no dependence on $r(t)$ for all $t$. In information theoretic terms, for the channel $\Lambda_B$ and the set of inputs above, this is equal to saying that $\chi(\Lambda_B(\mathcal{E}_t))=0$ for all $t$.\\
\\
{\bf RCQ Security} For RCQ protocols there is the additional condition of security. We say an authorised set $B$ is {\it secure} if the key generated by the protocol between dealer $d$ and players $B$ is perfectly secure. Note, that in order not to impose potentially impossible restrictions on $B$'s measurements, in these schemes a set $B$ is treated as one party, hence security is not guaranteed against cheaters within the set $B$. We expect such cheats can be overcome for all graph state schemes, but leave it to further work. As with all QKD schemes, an authenticated classical channel between $d$ and $B$ is required. Security will be shown against general attacks in a way which tolerates noise, as shown for a qudit extension of the Ekert protocol in \cite{SS10}. Previously security was only shown against intercept resend attacks \cite{HBB99,MS08,Keet10} (although to some extent cheating players within $B$ could be tolerated).

\bigskip

\section{Equivalence of QQ and RCQ} \label{SEC: equivalence}
We now explore the relationship between the existence of protocols for RCQ and QQ as described above.

For the QQ and RCQ schemes defined as above from a channel state $|CS\rangle$, with logical basis $\{|i_L\rangle\}$, the following relationships hold.
\begin{ppt}\ \label{PROP 123}\\
1. A QQ authorised set, is a RCQ authorised set.\\
2. A QQ unauthorised set is a RCQ unauthorised set.\\
3. A RCQ authorised set is a QQ authorised set.
\end{ppt}
Proof: $1$ and $2$ are clear since the access of the classical information is a special case of the quantum information. We directly deduce $3$ from the lemma $1$ of \cite{CW}, which says that
\begin{equation}
\chi(\Lambda_B(\mathcal{E}_0))+\chi(\Lambda_B(\mathcal{E}_1)) \leq I(\tau:\Lambda_B).
\label{bb}
\end{equation}

If a set $B$ can access in the RCQ protocol, after going through the associated quantum channel $\Lambda_B$, the classical information is accessible in at least two mutual unbiased bases $\{\ket{i}\}$ and $\{U\ket{i}\}$, this means that $\chi(\Lambda_B(\mathcal{E}_0))=\chi(\Lambda_B(\mathcal{E}_1))=\log(q)$, hence $I(\tau:\Lambda_B)\geq 2\log(q)$  Moreover from its definition we have $I(\tau:\Lambda_B)\leq 2\log(q)$. Hence $I(\tau:\Lambda_B)= 2\log(q)$, which means that the information is quantumly accessible. $\square$

We note that it is not true that a RCQ unauthorised set is automatically QQ unauthorised, for example the $(n,n)$ RCQ threshold schemes \cite{HBB99,MS08} are only $(n,0,n)$. However, as we will see additional mixing can address this and further, for pure state QQ the unauthorised sets exactly determined by the authorised sets, so that the connection between QQ and RCQ is exact.

\begin{figure}[h]
{\resizebox{!}{3cm}{\includegraphics{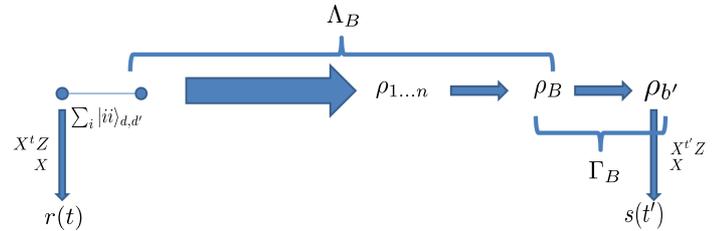}}} \caption{\label{FIG: RCQ} (Color online) Schematic of the RCQ scheme for the secure decoding. The channel state (\ref{Eqn: Channel State}) is generated by the dealer sending half an entangled state down the QQ encoding channel. The dealer then randomly chooses $t$ and measures in the associated basis, getting result $r(t)$. After tracing out of other systems to get $\rho_B$ authorised players $B$ perform QQ decoding map $\Gamma_B$, followed by a measurement associated with a random value $t'$, getting result $s(t')$. The strings  $r(t)$ and $s(t')$ are the raw strings from which the dealer and players $B$ can establish a secure random key using standard QKD techniques \cite{SS10}.}
\end{figure}

We will now show that a valid access structure for a RCQ protocol implies a secure key distribution.
\begin{ppt}A RCQ authorised set is a RCQ secure set.
\end{ppt}
Proof: From proposition \ref{PROP 123}.3 a RCQ authorised set is QQ authorised, hence there exists a decoding map $\Gamma_B$. Then we notice that the action of $\Gamma_B$ takes the channel state to a maximally entangled state between $d$ and $b'$.
To guarantee security we can define the measurements $\{M_B^{t'}\}_{t'=0}^{q}$ as first $B$ does $\Gamma_B$, then measures $\{X_{b'}^tZ_{b'}\}_{t=0}^{q-1}$ for $t \in \{0,...,q-1\}$ or $X_{b'}$ if $t=q$. For security, one can consider the step $\Gamma_B$ simply as part of the channel distributing the entangled state. The remaining part of the measurements coincide exactly with those in the extended six state protocol in \cite{SS10}, hence security follows directly from there. Note that the same connection holds if only two measurement settings were chosen, so in both directions two settings are sufficient to show equivalence. However, more settings can allow for better noise tolerance \cite{SS10}. See Fig.~\ref{FIG: RCQ}. Also, these measurements may not be the only ones allowing for a secure protocol. Indeed, the measurements in the RCQ schemes of \cite{MS08,Keet10} are local, and not of the form here, yet, the statistics can be shown to be equivalent and security is still guaranteed.

We summarize these results in the table \ref{TABLE}.
\begin{table}[!h] \label{TABLE}
\begin{tabular}{|c c c|}
\hline
RCQ & $\vline$ & QQ\\
\hline
$(n,k,k')$ & $\rightarrow$ & $(n,k,n-k)$\\
\hline
$(n,k,k'\geq n-k)$ & $\leftarrow$ & $(n,k,n-k)$\\
\hline
\end{tabular}
\caption{Relationships between RCQ and QQ protocols (Proposition 1).}
\end{table}

From these results we can immediately see that the schemes presented in \cite{CGL99} allowing for all QQ access structures can be used to give new RCQ protocols allowing for all access structures. Furthermore this can be done using high dimensional graph states \cite{MM12}.

We note again at this point that the equivalence presented here does not include all possible schemes for sharing classical secrets. This is clear since the equivalence presented also implies access structures violating no-cloning cannot work for RCQ schemes. In particular the use of QKD plus Shamir schemes does not prohibit access structures with more than one accessing set. Hence such schemes cannot be connected in a simple way to QQ schemes. Indeed this fact (as well as their possible intrinsic interest discussed at the end of this paper), is why we concentrate on RCQ schemes, so that we may make general, yet interesting statements of equivalence.

At this point we return to the question of dimensionality. In fact, with a small modification to considering the RCQ protocol for only two bases (the $t=0$ and $t=q$ bases), propositions 1 and 2 work for all composite dimension also. This follows from the proofs and the fact that the security in \cite{SS10}, and lemma (\ref{bb}), works for any dimension by restricting to these two bases.

\bigskip
\section{Connection to error correction} \label{SEC: QECC}

We now clarify the relationship between QQ , RCQ and quantum error correcting codes (QECC). A QECC encodes a space of dimension $\kappa$ onto $n$ systems (or shares), such that errors on some subsets of systems can be tolerated. A distance $d$  means that the code can tolerate the loss of $d-1$ shares (systems), or $(d-1)/2$ arbitrary errors at unknown locations. For  shares of dimension $q$ we denote a QECC as $((n,\kappa,d))_q$. Clearly one can use such an encoding as a QQ, and in the language of ramp schemes, if each share is a player, this means that $k = n-d+1$. Which players are unauthorised is apriori not given for a code and must be checked (see e.g. \cite{Gheo09,Gheo12}. Similarly it is clear that any QQ scheme is a QECC with $d=n-k+1$.

It was noticed in \cite{CGL99} that for the case of error correcting protocols encoding onto pure states, the situation becomes much simpler. It turns out that in this case it can be seen that the tolerance of a code to the loss of a set of shares $B_C$ is exactly equivalent to the same set $B_C$ not getting any information whatsoever about the encoded information. When used for QQ this means its ramp scheme parameter is $k' \leq n-k$. But by no cloning $k'\geq n-k$. Thus for all QQ with pure state encodings $k'=n-k$. For threshold schemes this reduces to $k=(n+1)/2$ as was explicitly stated in \cite{CGL99} (see also \cite{Gheo12} for linear codes and \cite{Sar12} for uses and applications to non-threshold schemes for qubits).

Furthermore, it gives a general relationship: a pure state QECC protocol $((n,\kappa,d))_q$ is equivalent to a QQ ramp scheme where all shares are considered as players with parameters $(k,k'=n-k,n)$. That is all such QECC are QQ ramp schemes with those parameters, and vice versa.

We can then ask what else is imposed by the relationship with error correction. One important question is that of share size. It can easily be seen that the Singleton bound implies that for (perfect) threshold schemes with pure state encoding $\kappa \leq q$. Hence, when $\kappa$ is a power of $q$, (as is the case for many codes, including all stabiliser codes), the only non trivial encoding satisfies $\kappa=q$ and all pure state (perfect and ideal) threshold schemes must be MDS codes (of dimension $1$) (see also \cite{RST05} for a rigourous information theorical based proof in both directions).
This implies something that has been shown for small $n$ cases in \cite{MS08}, which is that, for all pure state perfect threshold QQ secret sharing schemes encoding a secret equal to the size of each share (that is ideal schemes),  the dimension of each share must scale with $n$,
\begin{align} \label{eqn: q bound}
q\geq \sqrt{\frac{n+2}{2}}.
\end{align}
This bound, as explained in \cite{KKKS05}, follows from the fact that the code saturates the quantum Singleton bound. Moreover, the quantum MDS conjecture for such codes, also cited in \cite{KKKS05}, states that it would scale as badly as $q\geq \sqrt{n-1}$. This result extends the bounded maximal length given by theorem $6$ of \cite{Sar12} to qudit systems.

We note that the above results need only hold for pure state error correcting codes. The general schemes in this work have used pure state encoding. However, as mentioned earlier, mixed states encodings can also exist, though they will have purifications which can be phrased in our framework (hence in some sense they are also covered). It is interesting to consider what exactly our results mean for the mixed state schemes.

The first thing that we can say is that the relationships between QQ and RCQ will still hold in the mixed case. We have to be a bit careful by what we mean, but if we define both protocols in terms of the map $\Gamma_B$ from the original secret state $|\zeta\rangle_{d'}$ held by the dealer to the encoded version held by set of players $B$ (whereby QQ is a direct use of $\Gamma_B$ and RCQ is equivalent to the dealer sending half a maximally entangled state through $\Gamma_B$ then doing the measurements) and take the information theoretic definitions of authorised, unauthorised and secure given in section \ref{SEC: access}, the proofs for equivalence in section \ref{SEC: equivalence} follow through directly.

On the other hand, there do of course exist mixed state schemes which do not satisfy the error correction restrictions for pure state schemes above. Indeed, as pointed out in \cite{CGL99}, it is possible to go from $(k=n+1/2,n)$ to $(k,n-l)$ threshold schemes by throwing away $l$ systems. Clearly these mixed schemes to not satisfy $k'=n-k$. Such schemes were used in \cite{CGL99} to show that all QQ threshold schemes can be achieved using quantum Reed Solomon codes. It is these schemes which when translated to RCQ schemes (through our general relationship above) show all threshold (not violating no cloning) schemes are possible for RCQ also. Note also that this approach of discarding shares clearly holds in the RCQ extensions presented in this work, hence a QQ $(k,k',n)$ mixed state scheme implies a RCQ $(k,k',n)$ mixed state scheme.

Another set of schemes has been developed recently which do not satisfy the dimension restriction (\ref{eqn: q bound}) \cite{BCT,JMP,FG11,Gheo12}.
The idea of these schemes is to take pure state error correcting schemes, which are necessarily $(k,k'=n-k,n)$ ramp schemes, thus guaranteed quantum access to at least $k$, and add classical mixing on top to increase $k'$ arbitrarily (where classical information is distributed via classical secret sharing protocols over secure channels). Since the original quantum codes are no longer threshold schemes, they do not have to saturate the Singleton bound, and hence do not have to satisfy  (\ref{eqn: q bound}). However, even in this case it seems there are some restrictions on share size \cite{JMP}. Note also that both these sets of schemes can be purified, and their purifications clearly fall into our generalized schemes and must satisfy the above still, and although such purifications are impractical, this fact imposes restrictions on the mixed protocols also.

\bigskip
\section{Discussion} \label{SEC: discussion}
On the one hand, the error correction codes which were used to provide arbitrary access structures for QQ \cite{CGL99}, can, through the generalised scheme presented here, be used for RCQ, hence all access structures (not contracting the no cloning theorem) become possible. In addition we have seen that the mapping from QQ to RCQ allows for standard QKD security proofs to be used, implying full security within the authorised sets (where previously it was only known for limited attacks).

As was remarked in the introduction, applying  simple QKD plus existing classical secret sharing schemes
solves the same problem of an untrusted channel between the dealer and players as does RCQ. Nevertheless, we believe it interesting to study the existence of RCQ protocols in their own right - aside from the usefulness as a theoretical tool through the connection to QQ and QECC, they may be used as building blocks for more involved protocols. For example one may imagine using the redundancy of information present in RCQ in order to realise a more noise tolerant bipartite QKD - all the shares belong to one player Bob - so that Bob recover the information in the presence of noise (including erasure), i.e. a kind of quantum error corrected QKD. One may also imagine using RCQ as a means to authenticate a quantum channel using an authenticated classical channel - the only way that the correlations shared at the end of the RCQ protocol (that is, correlated measurement results), between the dealer and authorised set $B$, can be correct (i.e. close to equal) is if the quantum channel is close to perfect between the dealer and $B$ (indeed this is the essence behind the link to QQ). This could be used in combination with QQ as a way to test the channel and then use it for QQ for example. In this setting having both schemes using the same resources as presented here would make such combinations more practical in terms of both implementation and how they could be used together.

In the other direction, these results give a method for searching for error correcting codes starting from RCQ schemes. Checking the access structure (or error correcting capability) for RCQ can be more straightforward than checking the QQ case. We have seen that checking the access structure for only two bases suffices, for any dimensional system, to guarantee access (tolerance to loss) for quantum information too.  In particular for graph state schemes, many tools have recently been developed to phrase the conditions for secret sharing in solely graphical language, which have been used to search for new schemes \cite{Kashefi09,Gravier11,Gravier11b,MM12} which are therefore valid QQ and QECC schemes, and put bounds on the parameters that can be achieved. Through the general connection shown in this work, such techniques can also be used to search for quantum error correcting codes, in particular for higher dimensional codes which are seen to be necessary for the most efficient codes and general access structures.

\section{Acknowledgements}
We gratefully acknowledge discussions on the topics of this research with Simon Perdrix, Gilles Zemor and thank Vlad Gheorghiu for comments. We particularly thank Francesco Buscemi for discussions and pointing out reference \cite{CW} and Anthony Leverrier for explaining security proofs. We acknowledge financial support from ANR project FREQUENCY (ANR-09-BLAN=0410-03) and ANR Des program under contract ANR-08-EMER-003 (COCQ project).


\begin{thebibliography}{99}
\bibitem{NOTE}
We note that in \cite{MS08} and \cite{Keet10} these RQC schemes are referred to as CQ schemes - indeed RCQ are a subset of CQ schemes. We use this restricted notation to be clear that our results (and implied restrictions on possible access structures) do not apply to all possible CQ schemes.
\end{thebibliography}
\end{document}